\documentclass[%
 rsi,
 amsmath,amssymb,
 reprint,%
]{revtex4-1}

\usepackage{graphicx}
\usepackage{dcolumn}
\usepackage{bm}
\usepackage{color}

\usepackage[utf8]{inputenc}
\usepackage[T1]{fontenc}
\usepackage{mathptmx}

\begin{document}

\preprint{AIP/123-QED}

\title[Sample title]{Capacitive detection of magnetostriction, dielectric constant, and magneto-caloric effects\\ in pulsed magnetic fields}

\author{Atsushi Miyake}
 \email{miyake@issp.u-tokyo.ac.jp}
 \affiliation{The Institute for Solid State Physics (ISSP), The University of Tokyo, Kashiwa, Chiba 277-8581, Japan}

\author{Kenta Kimura}
\affiliation{Department of Advanced Materials Science, The University of Tokyo, Kashiwa, Chiba 277-8561, Japan}%

\author{Takumi Kihara}%
\affiliation{Institute for Materials Research, Tohoku University, Katahira, Sendai 980-8577, Japan}

\author{Hiroyuki Mitamura}
\affiliation{The Institute for Solid State Physics (ISSP), The University of Tokyo, Kashiwa, Chiba 277-8581, Japan}%

\author{Shiro Kawachi}
\email{Present address: Materials Research Center for Element Strategy, Tokyo Institute of Technology, Kanagawa 226-8503, Japan}
\affiliation{The Institute for Solid State Physics (ISSP), The University of Tokyo, Kashiwa, Chiba 277-8581, Japan}

\author{Tsuyoshi Kimura}
\affiliation{Department of Advanced Materials Science, The University of Tokyo, Kashiwa, Chiba 277-8561, Japan}

\author{Masashi Tokunaga}
\affiliation{The Institute for Solid State Physics (ISSP), The University of Tokyo, Kashiwa, Chiba 277-8581, Japan}%

\author{Makoto Tachibana}
\affiliation{National Institute for Materials Science, 1-1 Namiki, Tsukuba, Ibaraki 305-0044, Japan}

\date{\today}

\begin{abstract}

We report on the development of a capacitance measuring system which allows measurements of capacitance in pulsed magnetic fields up to 61~T. 
By using this system, magnetic-field responses of various physical quantities such magnetostriction, magnetic-field-induced change in complex dielectric constant, and magneto-caloric effect can be investigated in pulsed-magnetic-field conditions. 
Here, we examine the validity of our system for investigations of these magnetic-field-induced phenomena in pulse magnets. 
For the magnetostriction measurement, magnetostriction of a specimen can be measured through a change in the capacitance between two aligned electrodes glued on the specimen and a dilatometer. 
We demonstrate a precise detection of valley polarization in semimetallic bismuth through a magnetostriction signal with a resolution better than 10$^{-6}$ of the relative length change. 
For the magnetic-field-induced change in complex dielectric constant, we successfully observed clear dielectric anomalies accompanied by magnetic/magnetoelectric phase transitions in multiferroic Pb(TiO)Cu$_4$(PO$_4$)$_4$. 
For the measurement of magneto-caloric effect, a magnetic-field-induced change in sample temperature was verified for Gd$_3$Ga$_5$O$_{12}$ with a capacitance thermometer made of a non-magnetic ferroelectric compound KTa$_{1-x}$Nb$_x$O$_3$ ($x$ = 0.02) whose capacitance is nearly field-independent. 
These results show that our capacitance measuring system is a promising tool to study various magnetic-field-induced phenomena which have been difficult to detect in pulsed magnetic fields.

\end{abstract}

\maketitle

\section{\label{intro}Introduction}

Magnetic field is one of the essential external parameters to control physical properties that couple to various degrees of freedom, such as spin, orbital, and valence.
Extension of the applicable field region has led to the discoveries of new phenomena.
To generate magnetic fields over 50~T, pulse magnets are widely employed \cite{Herlach_1999}.
In addition to expanding the field region, the development of measuring techniques of physical quantities is indispensable.

Among various measurable quantities, we focus on the capacitance measurements.  
Capacitance measurements have been widely performed in static magnetic fields.
Through the capacitance measurements, one can investigate various physical quantities.
For example, a tiny lattice displacement can be detected through capacitance between two aligned electrodes: one of which attached to a specimen of interest is movable, while the other is fixed.
White developed measuring technique of thermal expansion, the relative length changes of the specimen as a function of temperature, using a capacitance dilatometer \cite{White_1961}.
By measuring capacitance changes as a function of magnetic fields, we can obtain the magnetostriction.
Furthermore, this capacitive method was extended for magnetic property measurements, for example, magnetic torque \cite{Griessen_1973} and magnetization \cite{Brooks_1987}. 
Capacitance is, of course, a direct measure of dielectric constant.
Besides, non-magnetic materials having large capacitance are used for temperature sensors, which is insensitive to the magnetic field.
Therefore, the capacitance measurement and its applications are useful for studies of the magnetic-field-induced phenomena that couple to multiple degrees of freedom in matter.

Although capacitance measurements are well established in static magnetic-fields, measurements in pulsed-magnetic fields have been rarely investigated up to now.
There are some difficulties for measurements in pulsed fields.
Usually, the conventional equipments used to measure capacitance, such as LCR meter, impedance analyzer, and automatic capacitance bridge, have rather long time constants required to obtain the capacitance.
Their time constants are typically several milliseconds, which is comparable to the time duration of our pulsed magnetic fields, e.g., 36~ms installed at ISSP, the University of Tokyo. 
These limitations motivate us to develop a capacitance measurement system in pulsed-magnetic-fields. 

In this paper, we describe our capacitance measurement system in pulsed-magnetic fields up to 61~T and some examples of results obtained using this system.
The capacitance measurements using a pulse magnet and a capacitance bridge are introduced in section \ref{capa}.
As a known application of capacitance measurements, we describe details of the capacitive detection of magnetostriction in \ref{MS_meas}.
\ref{PbCtanD} describes the results of complex dielectric measurements on a multiferroic material.
For further application of this system, we demonstrate magneto-caloric effect (MCE) measurement using a field-insensitive capacitance thermometer in \ref{MCE}.
Then, we conclude with a summary of our system and its applications.

\section{\label{capa}Capacitance measurements in pulsed-magnetic fields}

In this section, we introduce details of the capacitance measurement in pulsed-magnetic fields, which is a common technique through this manuscript. 

\begin{figure}[t]
\begin{center}
\includegraphics[width=\hsize]{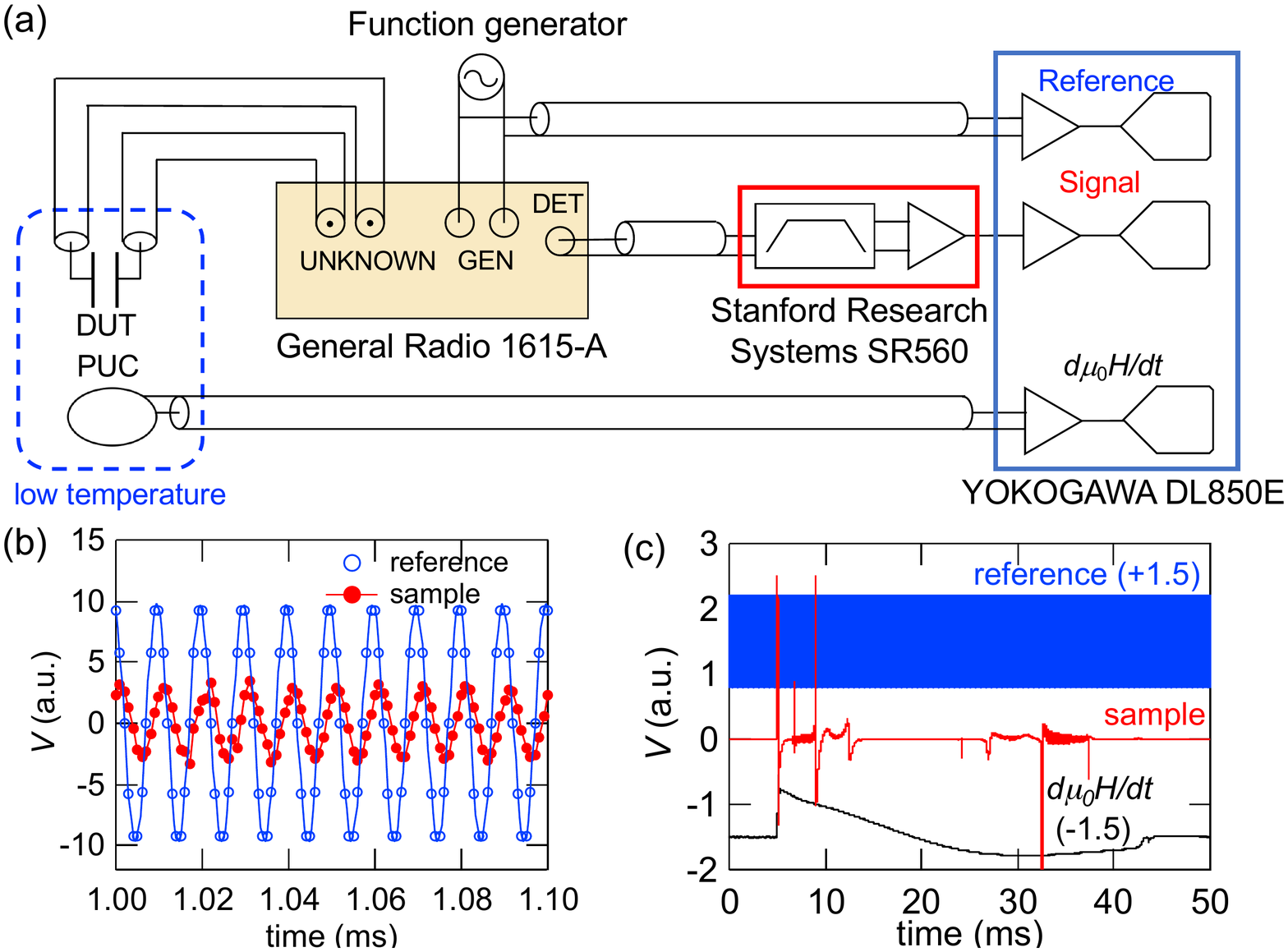}
\caption{(a) Schematic block diagram of the capacitance measuring system in pulsed-fields.
A sample (DUT) and a field pick-up coil (PUC) are located in a cryostat.
(b) An example of the reference and output (sample) signals as a function time recorded by the digitizer, which is a magnification of the pretrigger region of (c).
(c) An example of the acquired data, $d\mu_0H/dt$, sample, and reference signals as a function of time.
}
\label{raw}
\end{center}
\end{figure}

Capacitance was measured by using a capacitance bridge, General Radio (GR) 1615-A.
This bridge is a transformer-ratio type using a single decade of transformer voltage division and multiple fixed standard capacitors.
To measure the capacitance, we employed the three-terminal method.
AC voltage with a sinusoidal waveform generated by a function generator is used for exciting the bridge.
The output signal from the bridge [`DET' connection in Fig.~\ref{raw}(a)] is amplified by a preamplifier Stanford Research Systems SR560 passing through the band-pass filter.
The resultant signal is proportional to the off-balance between the sample, namely the device under test (DUT), and the reference standards in the bridge. 
The DUT is connected to the shielded connection through the coaxial cables.
The outer shield covers close to the DUT as possible (see Figs.~\ref{MS_cell} and \ref{setup}).

As shown in Fig.~\ref{raw}(a), the AC voltage output by a function generator is divided into two for correcting the phase shift caused by the use of the preamplifier and band-pass filter.
One is connected to the `generator' terminal of the bridge, and the other reference signal is directly connected to one channel of the digitizer, Yokogawa DL850E used here, being plugged a module 720268, having the 16-bit A/D converter.
The signals from the function generator (reference) and the amplified output (sample) are recorded simultaneously by DL850E in the single-shot acquisition mode.
The AC signal applying for the bridge is generated in burst mode to prevent change of phase position in each experiment.
The excitation amplitude is carefully selected not to distort the sample signal.

The signal from the sample is basically analyzed by performing a standard lock-in analysis numerically \cite{Machel_1995}: a sinusoidal signal is multiplied to the sample and reference signals, and then take a sliding average over multiple cycles of the sinusoidal wave to extract the real and imaginary components.
Figure~\ref{raw}(b) shows the reference and sample signals' raw data as a function of time at 0~T [magnification of the pretrigger region of Fig.~\ref{raw}(c)].
Here, the modulation frequency $f$ is 100~kHz, and the sampling rate is 1~MS/s.
A phase difference between the reference and sample signals can be clearly seen.
This phase shift is mainly caused by the use of a pre-amplifier and band-pass filter and gives extrinsic offsets in capacitance and conductance.
  
\begin{figure}[t]
\begin{center}
\includegraphics[width=\hsize]{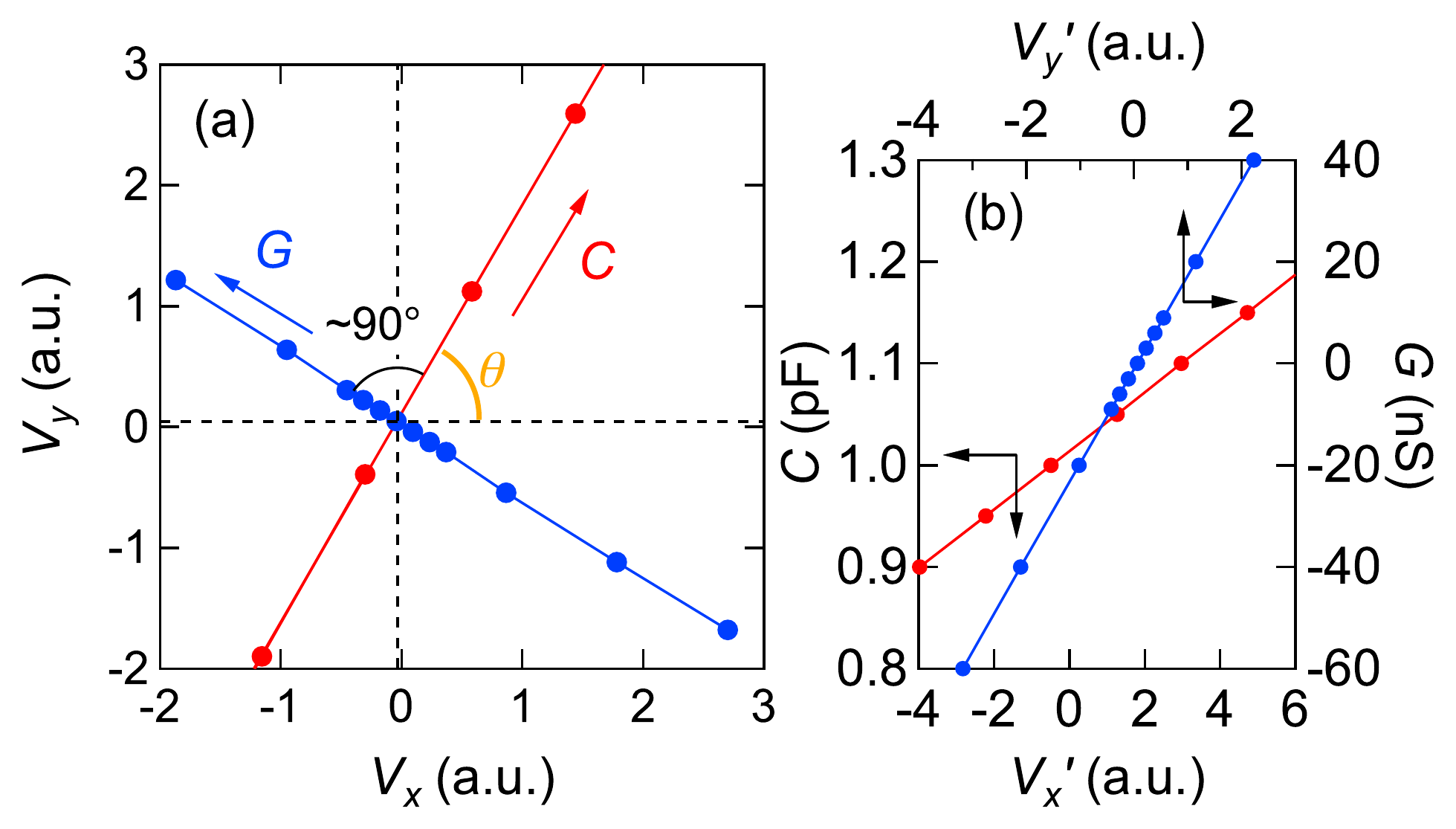}
\caption{(a) Reference capacitance and conductance dependences of the induced real ($V_x$) and imaginary ($V_y$) parts of demodulated output voltage in the bridge for Pb(TiO)Cu$_4$(PO$_4)_4$ at 4.2~K.
(b) The linear relation of capacitance (conductance) and the phase corrected complex voltage $V_x'$ ($V_y'$).}
\label{calib}
\end{center}
\end{figure}

We can determine the phase rotation $\theta$ between the sample and reference signals by comparing the real and imaginary parts of the demodulated sample signals.
Figure~\ref{calib}(a) shows an example of the real ($V_x$) and imaginary ($V_y$) parts of the demodulated signal with varying the capacitance ($C$) and loss component ($G$) references in the bridge.
With increasing the reference $C$ at constant $G$, $V_x$ and $V_y$ proportionally increase.
The $\theta$ is evaluated by $\arctan({V_y/V_x})$.
The change in the reference $G$ shows the similar trace rotated by 90~$^\circ$.

Figure~\ref{calib}(b) shows the calibration curves $C(V_x')$ and $G(V_y')$, where $V_x' (V_y')$ is the $\theta$-corrected demodulated sample signal $V_x (V_y)$.
We confirmed the clear proportionality between $C (G)$ and $V_x' (V_y')$.
We convert the $V_x' (V_y')$ to $C (G)$ that changes by applied magnetic fields using the slopes of these traces.

Pulsed-magnetic-fields up to 61~T, having $\sim 36$~ms time duration, were generated by pulse magnets installed at the International MegaGauss Science Laboratory in ISSP, at the University of Tokyo. 
The time evolution of the magnetic field is obtained by integrating the induced voltage in the pick-up coil [$d\mu_0H/dt$ signal in Fig.~\ref{raw}(c)].
The measuring probe was mainly made of non-metallic and non-magnetic materials, namely to reduce electromagnetic force and heating during the magnetic-field pulse (Figs.~\ref{MS_cell} and \ref{setup}).
The sample and wires are rigidly fixed by varnish or epoxy resin, to minimize the voltage noise caused by mechanical vibration.  

\section{\label{MS_meas}Magnetostriction measurements using a capacitive method in pulsed-field}

In this section, we discuss details of the magnetostriction measurement in pulsed-magnetic fields using a capacitance dilatometer.
First, we briefly review the various techniques of magnetostriction measurement in pulsed-magnetic field.
Next, the details of our dilatometer assembly are introduced.
Finally, experimental results of linear magnetostriction obtained with a Bi sample are presented.

\subsection{\label{MS_review}Various ways to measure magnetostriction in pulsed-magnetic field}

The linear (volume) magnetostriction, i.e., the change of sample length $\Delta L$ (volume) as a function of a magnetic field, is a thermodynamic constant and includes fruitful information of the matters.
Magnetostriction measurements in pulsed-magnetic fields have been developed utilizing various techniques used in static fields as follows.

The strain-gauge method is one of well-known techniques to measure the change in the sample length.
The resistive gauge used in the static field was adopted for the studies in pulsed-fields \cite{Ricodeau_1972}. 
The change in resistance of gauge bonded to the sample of interest is proportional to the change in the sample length.
However, the application of too large current to obtain a high-resolution result heats the sample due to the Joule heating in the gauge \cite{Knafo_PC}.

To eliminate the heating, piezoelectric materials are also used as the strain gauges \cite{Levitin_1992, Ding_2018}.
Magnetostriction measurement is attainable from the induced voltage due to the piezoelectric effect caused by the change in sample length.
Because the voltage is self-induced, an external power source is unnecessary, and thus no Joule-heating of the gauge emerges.
The use of these simple gauges is advantageous to measure the field-angular dependence of magnetostriction \cite{Ding_2018}.

Recently, an optical method using a fiber Bragg grating (FBG) has been developed for use in the pulsed-magnetic field by Daou {\it et al} \cite{Daou_2010}.
Since the use of optics, it has a significant advantage of escaping from the inherent electromagnetic noises from the pulsed-field generation.
Further developments to use even under ultrahigh magnetic fields exceeding 100~T (using a destructive pulse magnets) have attained \cite{Ikeda_2017, Jaime_2017, Ikeda_2018}.
One can also find a successful work to 100.75~T using a nondestructive pulse magnet \cite{Jaime_2010}.
The resultant accuracy is the order of $10^{-7}$ in pulsed-fields \cite{Daou_2010, Jaime_2017}.
In addition to longitudinal magnetostriction measurement, transverse magnetostriction measurements ($\Delta L~\perp~\mu_0H$) were also reported \cite{Rotter_2014, Radtke_2015}.

The capacitance method is also a well-known alternative technique for magnetostriction measurement in the static-magnetic fields.
Kido achieved the adaptation to use it in the pulse-magnetic field using a small nonmagnetic and nonmetallic dilatometer \cite{Kido_1989}.
A simultaneous measurement of magnetostriction and magnetization in the pulsed magnetic-field was also developed \cite{Doerr_2008}.

Here, we introduce the longitudinal magnetostriction measurement using the capacitance method in more detail.
Although we will not discuss here, this method can be adapted for the transverse magnetostriction measurements \cite{Kawachi_2017}.

\subsection{\label{MS_cell_detail}The detail of capacitance dilatometer assembly}

\begin{figure}[htbp]
\begin{center}
\includegraphics[width=\hsize]{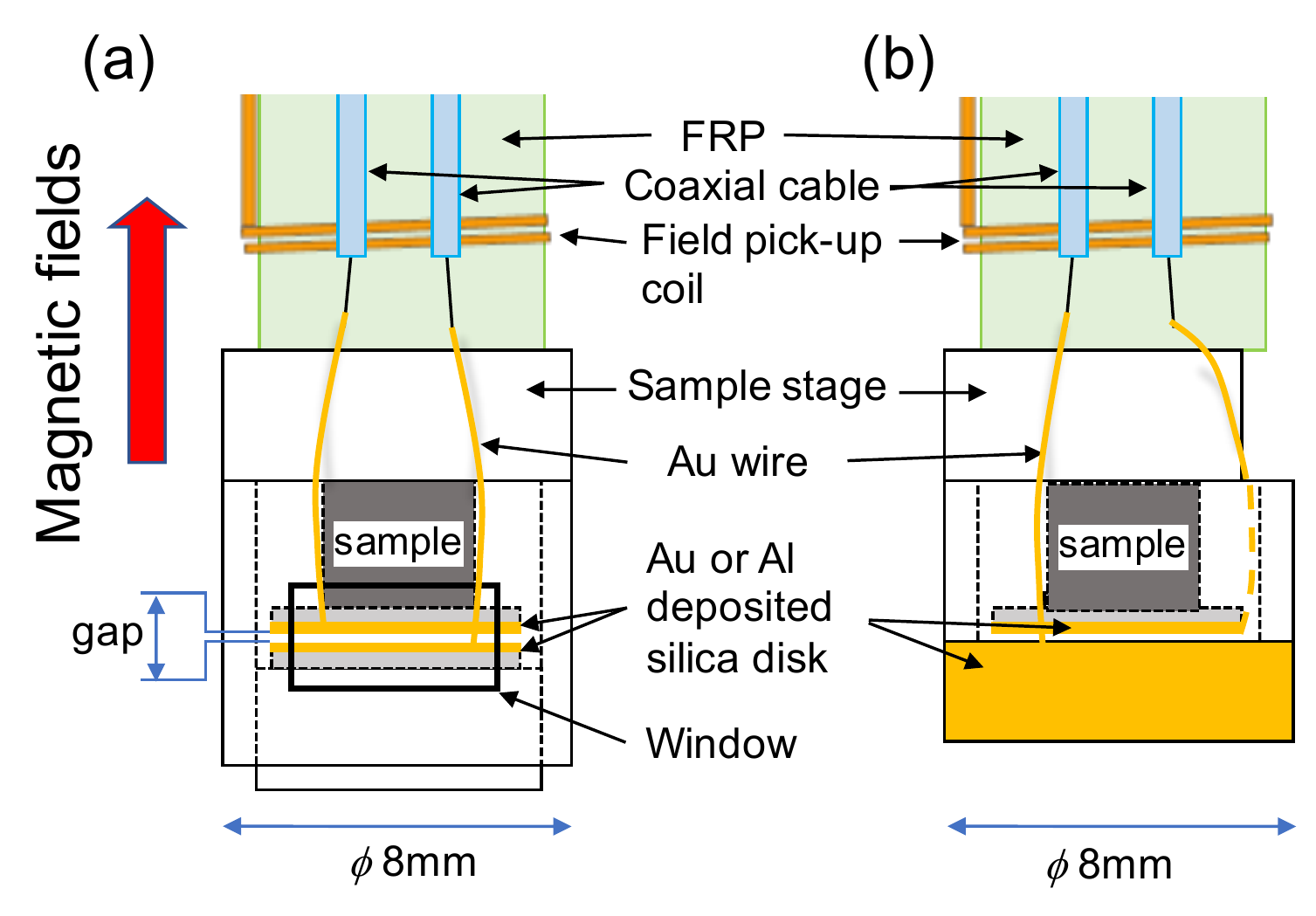}
\caption{Schematic drawings of two types of  of our capacitance dilatometers for the longitudinal linear magnetostriction measurement.
(a) Piston-cylinder type.
The rectangular of the center part of the cell (solid thick line) is a window to access the electrodes of condenser.  
(b) Fixed-electrodes type.
To make a contact with the electrode on the sample, a side of the cylindrical sample stage is flattened.} 
\label{MS_cell}
\end{center}
\end{figure}

For the capacitive magnetostriction measurements, we use a homemade capacitance dilatometer cell.
Figure~\ref{MS_cell} draws a schematic of our dilatometers used for longitudinal linear magnetostriction measurement.
The design of these cells is quite similar to those reported in \cite{Kido_1989, Doerr_2008}.
The cells are made of nonmetallic and nonmagnetic materials, such as the machinable ceramic Macor and fused silica.
We change the length of the cells depending on the sample length along the magnetic field. 
The diameter of the cell used here is typically 8~mm.
The maximum available diameter is limited by the sample space of the cryostat.

The cell consists of two electrodes, a cylinder, and two stages, which are assembled at room temperature, as shown in Fig.~\ref{MS_cell}(a).
To glue them, we usually use stycast 1266 or varnish and keep these materials in a vise during the hardening of the glue.
The construction procedures are as follows.
First, the sample is polished carefully to have a parallel flat surface on each side along the magnetic field direction, and then glued to the sample stage.
Second, an electrode is glued on the top of the sample.
The electrode is an Au or Al-deposited disk made of fused silica.
For depositing Au, Ge is deposited as a buffer between Au and the silica.
The typical thickness of the metal film is 150~nm. 
Using the metal film as the capacitance electrodes, eddy-current heating on the electrodes discussed in Ref.~\cite{Doerr_2008} is not observed in our system.
Third, another electrode is fixed to the stage.
As shown in Fig.~\ref{MS_cell}(a), this stage can move like a piston for the cylindrical cell and can adjust a gap, namely the distance between capacitance electrodes.
Fourth, we make a gap between electrodes.
The gap distance should be small enough, since the gap distance is inversely proportional to the capacitance.
To make the gap while keeping the electrodes parallel, we sandwich a polyimide film with a typical thickness of 5-25~$\mu$m between the electrodes through a window of the side of the cell during gluing [Fig.~\ref{MS_cell}(a)].
Here, we pay attention to the effect of thermal contraction to determine the initial gap size, as will be discussed later.
Finally, the film is removed after the hardening of the glue.

Another cylinder-type cell is also shown in Fig.~\ref{MS_cell}(b).
Compared to the previous piston-cylinder type configuration, an electrode is directly glued to the bottom of the cylinder. 
Therefore, the diameter of the fixed electrode is larger than that on the sample.
Thus, the displacement along the radius direction of the cell is not necessary to be considered, which is an advantage compared to the piston-cylinder type.
The hight of the cylinder determines the gap distance. 
And hence, the hight should be tuned to make a suitable gap distance.
Using this type of dilatometer of fused silica, we can also measure the temperature dependence of $\Delta L/L$, namely thermal expansion, which will be demonstrated later.

The magnetic-field variation of the capacitance between electrodes of condenser can estimate magnetostriction with the following relation.
 
\begin{equation}
\frac{\Delta L}{L}=\frac{\epsilon_0 A}{L}\left(\frac{1}{C_0}-\frac{1}{C}\right),
\label{MS}
\end{equation}
where $\epsilon_0$, $A$, $C_0$, and $C$ are the permittivity of vacuum $\epsilon_0=8.854\times 10^{-12}$~F/m, cross-section of electrode, the initial capacitance, and capacitance measured as a function of field.
Since $C$ is proportional to $A$, we select the electrode disk diameter to 6~mm here to be as large as possible to match the cell.
It is also noted that the $C_0$ and $C$ include the stray capacitance of the circuit.
With increasing the capacitance by reducing the gap and increasing the electrode area, the parasitic effect becomes negligibly small.  

\begin{figure}[htbp]
\begin{center}
\includegraphics[width=\hsize]{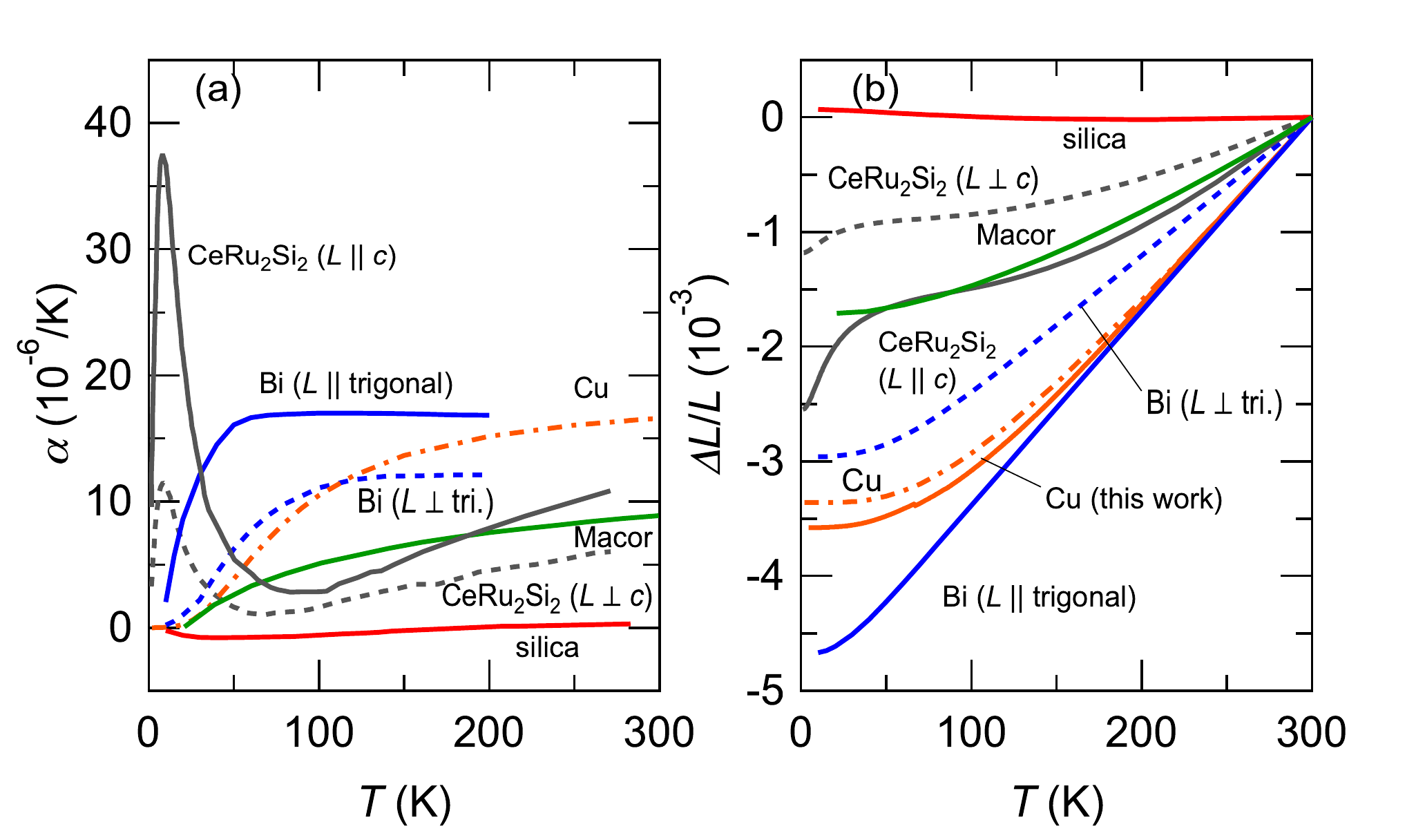}
\caption{Temperature dependences of (a) linear thermal expansion coefficient, $\alpha \equiv (1/L)(dL/dT)$, and (b) linear thermal expansion of various materials.
Here, $\Delta L$ is measured from 300~K. 
Thermal expansion of Cu measured by using the dilatometer of Fig.~\ref{MS_cell}(b) is shown for comparison.
The data of Macor, fused silica (Vitresoil 1400), Bi, Cu, and CeRu$_2$Si$_2$ are taken from Refs. \cite{Lawless_1979}, \cite{White_1973}, \cite{Bunton_1969}, \cite{White_1974}, and \cite{Lacerda_1989}, respectively.} 
\label{TE}
\end{center}
\end{figure}

To obtain high accuracy, we have to pay sufficient attention to the temperature dependence of the gap caused by thermal contraction.
To estimate the suitable gap distance set at room temperature, we compare the thermal expansion of various materials reported previously in Fig.~\ref{TE}. 
In our case, the cell consists of Macor or fused silica.
Although Macor has a relatively small thermal expansion coefficient $\alpha \equiv (1/L)(dL/dT)$, the coefficient is not negligible for some materials \cite{Lawless_1979}.
The $\alpha$ of Macor, for example, is comparable to or smaller than those of a heavy-fermion paramagnet CeRu$_2$Si$_2$, whose $\alpha$ is anisotropic reflecting the tetragonal crystal structure \cite{Lacerda_1989}.
In contrast to Macor, the $\alpha$ of the silica is much smaller among the materials compared in Fig.~\ref{TE}(a), although this coefficient varies with the temperature \cite{White_1973}.
The length change between room temperature and low temperature is negligibly smaller than those of the other materials, as seen in Fig.~\ref{TE}(b).
Using a dilatometer made of the silica, we also demonstrate the thermal expansion measurement of Cu [Fig.~\ref{TE}(b)]. 
The result agrees with the previous report within 10~\% \cite{White_1974}.
A number of the thermal expansion data for other materials can be also found in the textbook \cite{Baron_1999}.

\subsection{\label{Bi_MS}Example of magnetostriction measurement: the case for the valley polarization in Bi}
 
As an example of the longitudinal magnetostriction measurements using our capacitance dilatometer and measuring system, experimental results obtained with a Bi sample are presented.
We selected this sample, because its magnetostriction was already investigated up to 40~T \cite{Michenaud_1982, Iye_1983}.

Bi is a nearly compensated semimetal having an almost equal number of electrons and holes with typical densities of $10^{17}$~cm$^{-3}$ (see, for example \cite{Fuseya_2015}). 
The Fermi surfaces consist of highly anisotropic one hole pocket and three electron pockets.
The multiple constant energy landscapes at the different positions in the momentum space are called valleys.
The control of the valley is essential in terms of not only fundamental physics but also for applications.
Application of magnetic fields lifts the degeneracy of the valleys and causes complete valley polarization at high fields \cite{Zhu_2017}, although several models predict different scenarios \cite{Zhu_2012}.
Magnetostriction measurements may provide us the thermodynamic evidence for this valley polarization.

Here, we demonstrate high-resolution longitudinal magnetostriction measurement of Bi, using our systems.
A 9.0-mm-long sample was set as drawn in Fig.~\ref{MS_cell}(a) with the gap of $\sim7~\mu$m at room temperature evaluated from the capacitance value.
The gap increases to $\sim19~\mu$m at 1.4~K owing to the thermal contraction.
In this magnetostriction measurement, we applied large excitation voltages of 90 V$_{\rm p-p}$ using a power amplifier (Turtle Industry Co. Ltd., T-HVA02) to improve the resolution better than $10^{-6}$.
The sample signal was first amplified by a differential preamplifier (NF Corporation, SA-400F3) and filtered by a band-pass filter.
The measuring frequency is 50~kHz.

\begin{figure}[htbp]
\begin{center}
\includegraphics[width=\hsize]{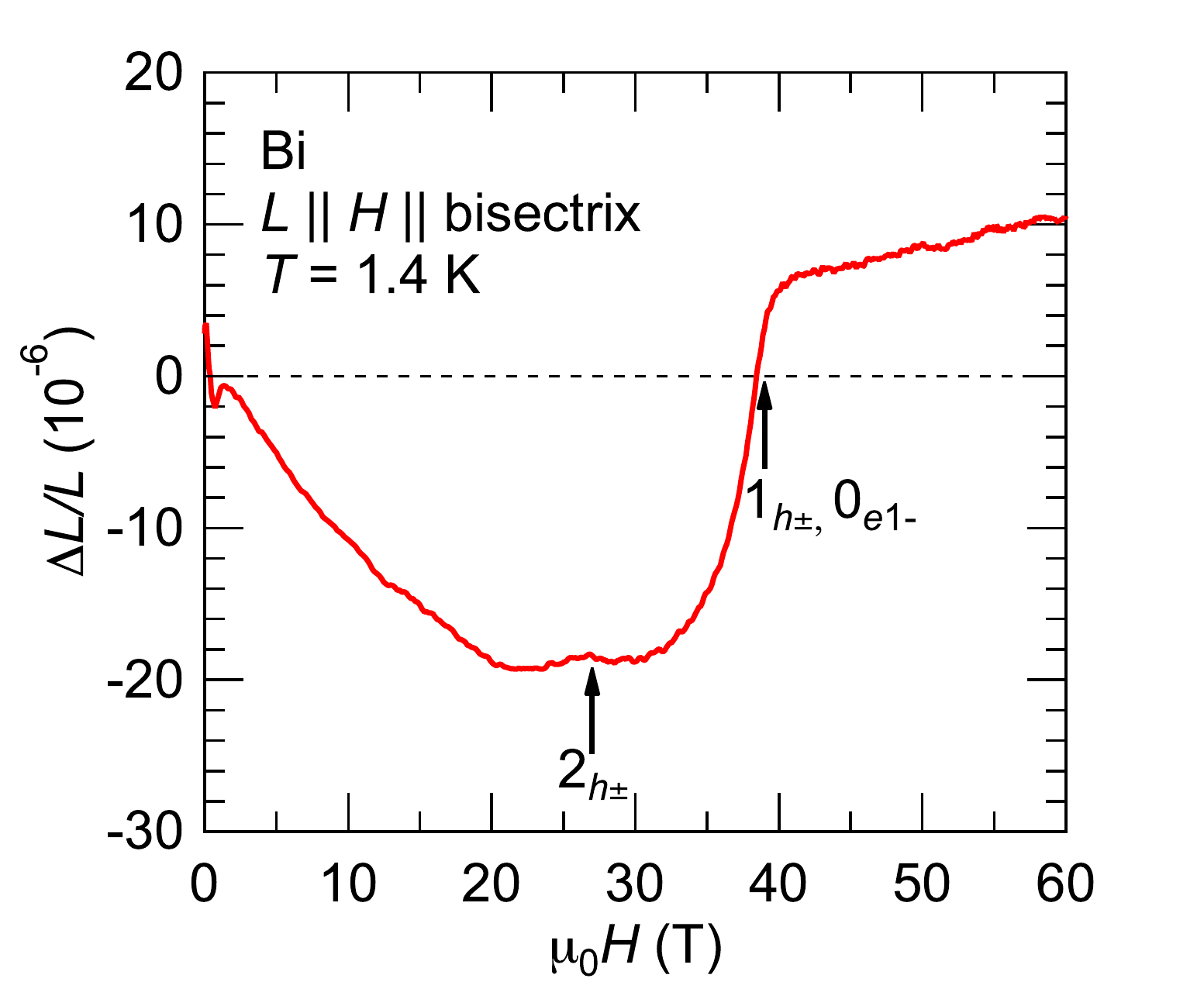}
\caption{Magnetic field dependence of the longitudinal magnetostriction $\Delta L/L$ in Bi for $H~||~$bisectrix axis at 1.4~K.
The horizontal dashed line is the zero magnetostriction.
Arrows indicate the fields where the Landau level crosses with the Fermi level.
}
\label{Bi}
\end{center}
\end{figure}

Figure~\ref{Bi} shows the longitudinal magnetostriction in Bi for $H~||~$bisectrix axis at 1.4~K.
A kink at 27~T and a sign-change at 39~T (marked by arrows) are clearly observed, which correspond to the quantum oscillation of the hole Landau subband and valley polarization, respectively, as discussed in \cite{Iwasa_2019}.

This capacitance method has an advantage over the strain gauge method because Joule heating is negligible during measurements.
In addition, the small cell size of this capacitance method makes it possible to change the magnetic field orientation while preserving the setup \cite{Kawachi_2017}.
However, for the materials with an unknown thermal expansion coefficient, we sometimes have to repeat the setups to obtain the optimal gap size.
Also, imperfect alignment of the electrodes leads to the uncertainty of the
absolute value.
Therefore, the disadvantage is that the smaller the sample, the more difficult it is to set up.

\section{\label{PbCtanD}Example of dielectric constant measurement: the case for Pb(TiO)Cu$_4$(PO$_4)_4$}

In the study of multiferroic materials, electric polarization measurements are powerful tools to investigate phase transitions induced by magnetic fields \cite{Tokura_2014}. 
Since higher sweeping rate of magnetic fields yields larger displacement current whose time-integral corresponds to magnetic-field-induced electric polarization, measurements using pulsed-magnetic-fields have great advantage in the study of multiferroic materials {\cite{Popov_1993}.
Therefore, a number of studies using pulsed-fields on magnetic-field-induced electric polarization in multiferroic materials have been reported to date.}
This technique works only when the transition involves a change in a macroscopic electric polarization. 
Dielectric constant, on the other hand, can detect various types of dielectric property changes, including antiferroelectric transitions and non-(anti)ferroelectric structural transitions. 

The dielectric constant $\epsilon$ is simply obtained by capacitance measurements through the relation,
\begin{equation}
 \epsilon = \frac{Cd}{A},
 \label{diele}
 \end{equation}
where $C$, $d$, and $A$ are the capacitance, distance between the electrodes on the sample, and the cross-section area of the electrode for the sandwich-type condenser configuration.
However, dielectric constant measurements using a pulse magnet have been rarely reported so far \cite{Zapf_2010, Chen_2019}. 
This motivates us to establish the capacitance-based dielectric constant measurement system suitable for pulsed-magnetic-field experiments.

\begin{figure}[htbp]
\begin{center}
\includegraphics[width=\hsize]{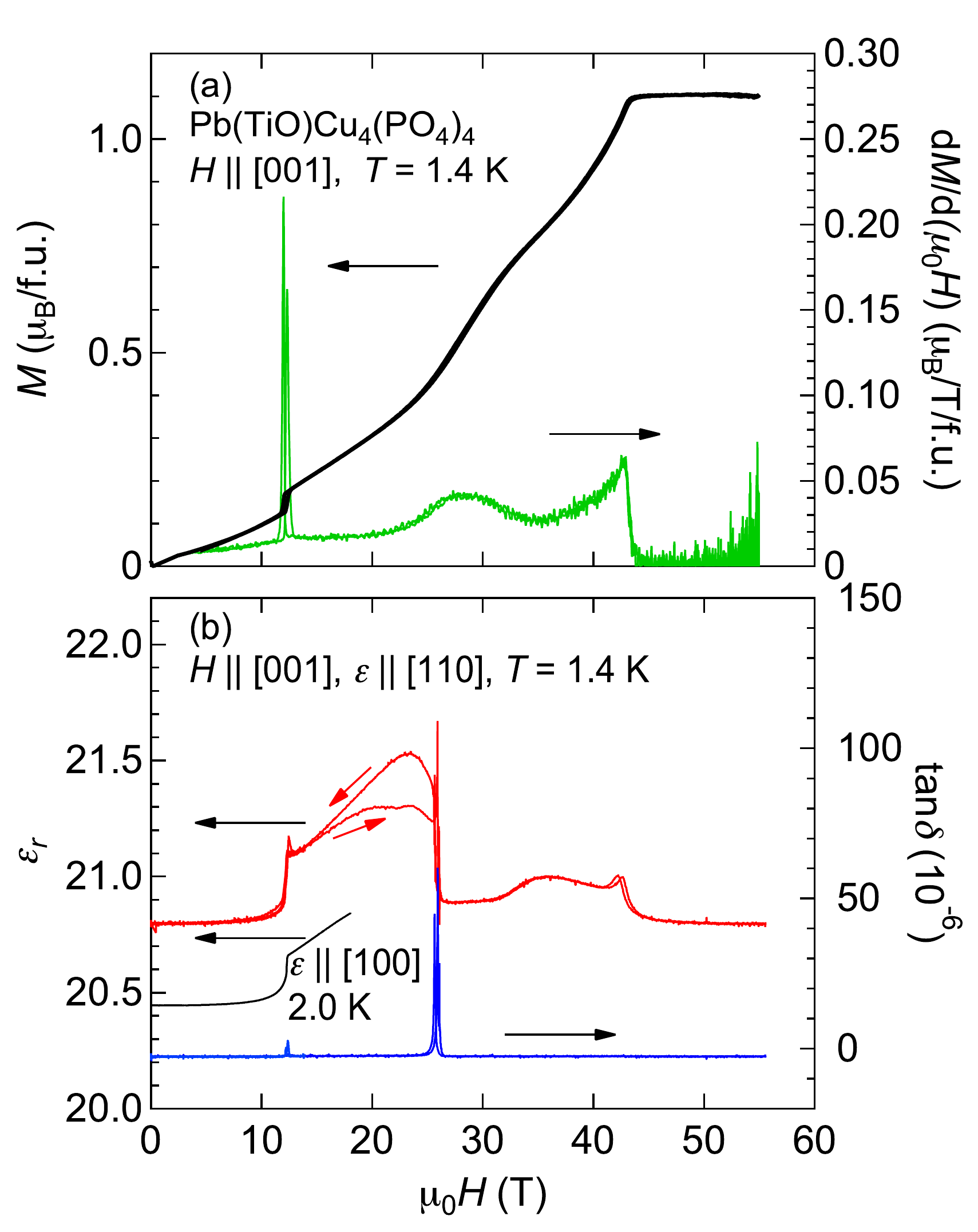}
\caption{(a) Magnetic field dependence of magnetization (left axis) and the differential susceptibility $dM/d(\mu_0H)$ (right axis) of Pb(TiO)Cu$_4$(PO$_4)_4$ for $H~||~[001]$ at 1.4~K taken from Ref.~\cite{Kimura_2018}.
(b) Magnetic field dependence of dielectric constants $\epsilon_r$ (left axis) and $\tan{\delta}$ (right axis) along the [110] direction for $H~||~[001]$ at 1.4~K.
For comparison, $\epsilon_r$ along the [100] direction for $H~||~[001]$ at 2.0~K measured in static field, taken from \cite{Kimura_2019}.
The data is offset by +0.5.}
\label{Pb_C}
\end{center}
\end{figure}

To demonstrate the performance of our measurement system, we select multiferroic Pb(TiO)Cu$_4$(PO$_4)_4$ (abbreviated as PbTCPO hereafter) as a test material, which shows several magnetic anomalies in high magnetic fields.
Single crystals of PbTCPO were grown by the slow-cooling method described in Ref~\cite{Kimura_2018B}.
Figure~\ref{Pb_C}(a) shows the previously reported magnetic-field dependence of magnetization for $H~||~c$ at 1.4~K in PbTCPO \cite{Kimura_2018}. 
The magnetization shows a clear jump at $\mu_0H_c$~=~12.3~T, hump structure at around 30~T, and spin saturation at $\sim 42$T. 

Figure~\ref{Pb_C}(b) shows the magnetic field dependence of the complex dielectric constant along [110] in PbTCPO at 1.4~K.
The $\epsilon_r$ is a specific dielectric constant expressed as $\epsilon_r = \epsilon/\epsilon_0$.
The $\tan{\delta}$ is a loss tangent defined as $\tan{\delta} = \frac{G}{2\pi fC}$.
The previous experiments in a steady magnetic field reported that this metamagnetic transition accompanies on antiferroelectric-like transition \cite{Kimura_2019}: an anomaly appears not in the electric polarization but in the $\epsilon_r$ in the tetragonal $c$ plane, as shown in Fig.~\ref{Pb_C}(b). 
As shown in this figure, the data obtained in a pulsed field reasonably reproduces that in a steady field.
In a pulsed field, we can identify the anomaly in $\epsilon_r$ corresponding to the spin saturation at $\sim 42$~T.
Moreover, we observed a steep change in $\epsilon_r$ and $\tan\delta$ at around 26~T, which corresponds to a broad hump anomaly in magnetization.
Although the origin is unclear yet, our dielectric measurement detects this anomaly.

Another example demonstrating the dielectric constant measurements using our system was already published for the multiferroic system Co$_2$V$_2$O$_7$ \cite{Chen_2019}.
Clear peaks are observed in dielectric constant when the system crosses the ferroelectric phase with the field scan.

We believe that the dielectric constant measurement is susceptible to detect some changes of magnetic structures.
In addition to the widely studied magnetization and electric polarization in pulsed-magnetic-fields, the dielectric constant is a mighty probe for magnetic insulators, such as multiferroic and quantum spin systems. 
Since it is rarely studied the dielectric constant beyond 50~T in many systems interested, our technique may accelerate more in-depth understanding.

\section{\label{MCE}Magneto-caloric effect measurement in pulsed-magnetic fields using a capacitance thermometer}

In the previous section, we demonstrated precise measurements of complex dielectric constants in pulsed fields.
We can develop this technique to magneto-caloric effect (MCE) measurements using a nonmagnetic capacitance thermometer.
First, we briefly touch upon some examples of MCE measurements in pulsed-magnetic fields, and mention some problems to be overcome.
Next, we discuss the usability of a non-magnetic ferroelectric KTa$_{1-x}$Nb$_x$O$_3$ as a magnetic-field-insensitive capacitance thermometer.
Finally, we demonstrate the MCE measurement of a magnetic refrigerant Gd$_3$Ga$_5$O$_{12}$ using the above thermometer in pulsed-magnetic field.

\subsection{\label{MCE_meas} MCE measurements in pulsed-magnetic field}

MCE, a magnetic-field-induced change in temperature of the specimen in adiabatic conditions, is quite attractive not only for the fundamental investigation of physical properties but for applications of magnetic refrigerators (see, for example \cite{Gschneidner_2005}).
The MCE measurements at low temperatures are very efficient at investigating phase transitions, since the entropy changes detected by MCE reflect changes in any order parameters.
In addition, the magnetic Gr$\ddot{\rm u}$neisen parameter, which can be determined by MCE, has recently received much attention in the study of physics in the quantum critical point \cite{Zhu_2003, Garst_2005}.

The experimental techniques to investigate the MCE by field-sweep using a pulse magnet may be classified into two types.
One is an indirect method.
The MCE is evaluated from thermodynamic relation by comparing the magnetization data measured in an adiabatic condition to the calculated isothermal one \cite{Levitin_1997}. 
The difference between them arises from the MCE when the calculation reproduces the isothermal magnetization. 
The other one is more direct measurement by sensing the temperature change of material in adiabatic condition as a function of the magnetic field \cite{Dankov_1997, Kohama_2010, Kihara_2013, Kohama_2013, Gottschall_2016}.
Because of the short duration of the  pulsed-magnetic field, the adiabatic condition can be generated rather easily.
To monitor the temperature, thermocouple \cite{Dankov_1997, Gottschall_2016} or small resistance thermometer \cite{Kohama_2010, Kihara_2013, Kohama_2013} has been used.

In this direct method, compensation for the magneto-Seebeck and magnetoresistance effect of the sensor is essential.
A commercially available Cernox\texttrademark  ~(zirconium oxynitride) thermometer, for example, is known to show significant non-monotonic magnetoresistance \cite{Brandt_1999, Scott_2003}, in particular at temperatures below 2~K \cite{Brandt_1999, Rosenbaum_2001, Kohama_2010}.
Moreover, we have to pay attention to the Joule heating in the resistance measurements of the thermometer at low temperatures.

A better choice would be a capacitance thermometer \cite{Murphy_2001, Tetsuka_2006, Shimada_2006}.
The advantages of using capacitance thermometers are their small magnetic field dependence of capacitance (magnetocapacitance) and no self-heating.
For example, a magnetocapacitance effect of 0.13\% at 90~mK and 19~T was reported for SrTiO$_3$ \cite{Naughton_1983}.  
This value is negligible compared to the temperature coefficient.
Therefore, it is believed that the temperature dependence of capacitance measured at zero magnetic field can be adopted for finite magnetic fields. 
In order to use capacitance thermometer at higher fields, we have to clarify the magnetocapacitance under such conditions.

\subsection{\label{KTN}Temperature measurement of capacitance thermometer in pulsed fields}

\begin{figure}[htbp]
\begin{center}
\includegraphics[width=\hsize]{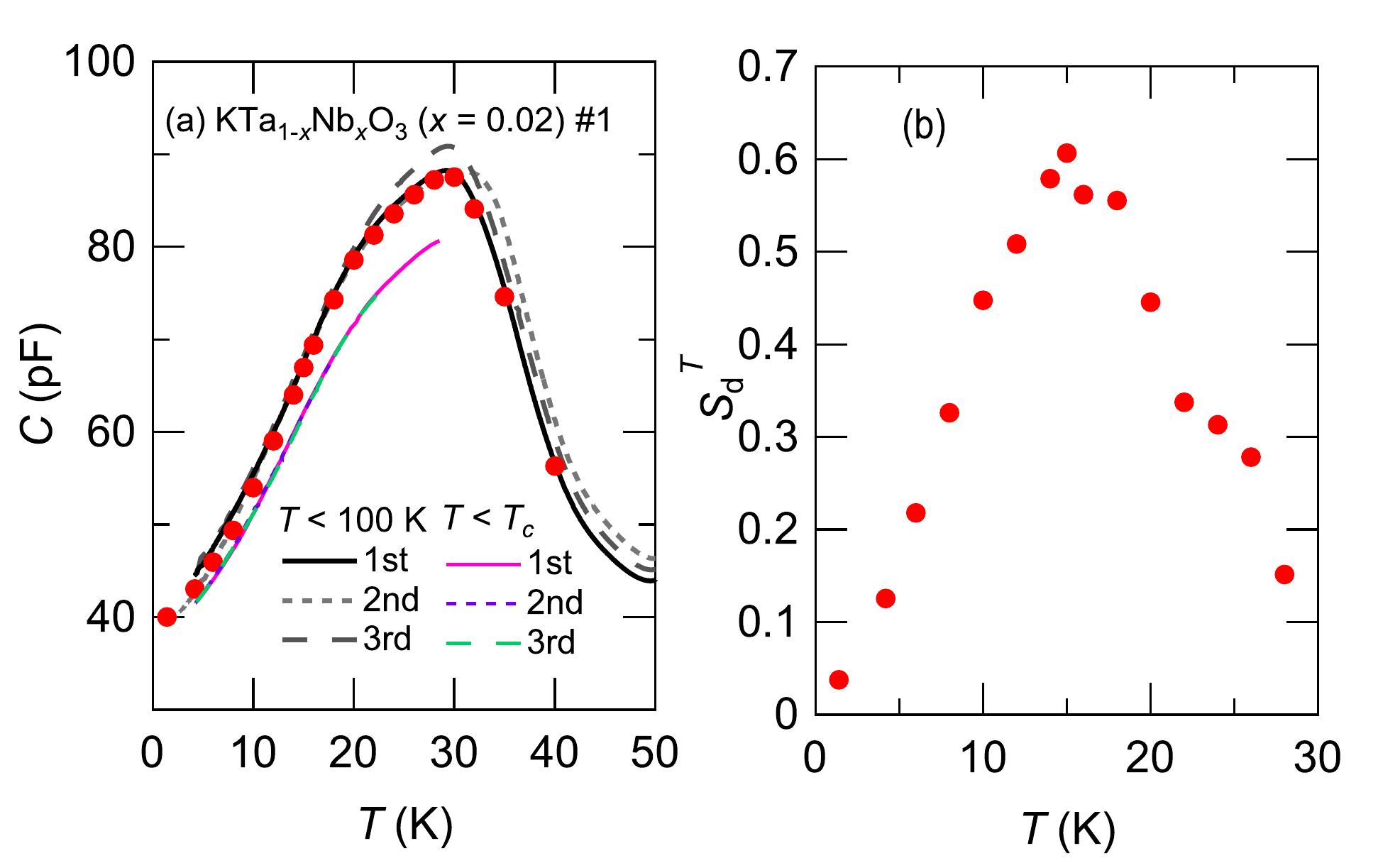}
\caption{
(a) Temperature dependence of capacitance of KTa$_{1-x}$Nb$_x$O$_3$ ($x$~=~0.02) \#1 at zero magnetic fields measured at 1~kHz (lines) and  50~kHz (closed circles).
The thermal cycle dependences obtained at 1~kHz for $T<$~100~K and $T<$~30~K are also shown, which are measured as temperature increases. 
(b) Temperature dependence of the dimensionless temperature sensitivity $S_d^T = (T/C)(dC/dT)$ by using the $C(T)$ data measured at 50~kHz.}
\label{CH}
\end{center}
\end{figure}

To measure the temperature change in pulsed-magnetic fields, we utilize a nonmagnetic-ferroelectric material KTa$_{1-x}$Nb$_x$O$_3$ (abbreviated as KTN) as a capacitance thermometer.
The ferroelectric transition temperature $T_c$ of KTN is known to be tuned by Nb-doping to the Ta site \cite{Rytz_1983}.
Single crystals of KTN with nominal $x$~=~0.02 were prepared by the flux methods \cite{Tachibana_2015}.
To characterize our KTN samples, we measured the temperature dependence of capacitance at zero magnetic field using the Andeen-Hagerling (AH) 2500A automatic bridge with a measuring frequency of 1~kHz and GR 1615A with 50~kHz.
Here, we limit the excitation voltage to avoid non-linear $P-E$ response.
As shown in Fig.~\ref{CH}(a), capacitance of KTN \#1 displays a peak at $T_c\sim$~30~K, which corresponds to the ferroelectric transition reported previously \cite{Rytz_1983}.
On cooling below $T_c$, capacitance monotonically decreases with a significant temperature coefficient.
This simple temperature dependence below $T_c$ is suitable for use as a thermometer.

To evaluate the feasibility of our capacitance thermometer, we introduce the dimensionless sensitivity $S_d^T = (T/C)(dC/dT)$.  
As shown in Fig.~\ref{CH}(b), $S_d^T$ varies approximately from 0.04 (1.4~K) to 0.61 (15~K).
These values in this temperature region are slightly larger than those of other ferroelectric perovskites, such as Ba$_x$Sr$_{1-x}$TiO$_3$ \cite{Tetsuka_2006} and Sr$_{0.95}$Ca$_{0.05}$TiO$_3$ \cite{Shimada_2006}.  
These facts confirm that our KTN sample is suitable for the use of capacitance thermometer.

\begin{figure}[htbp]
\begin{center}
\includegraphics[width=\hsize]{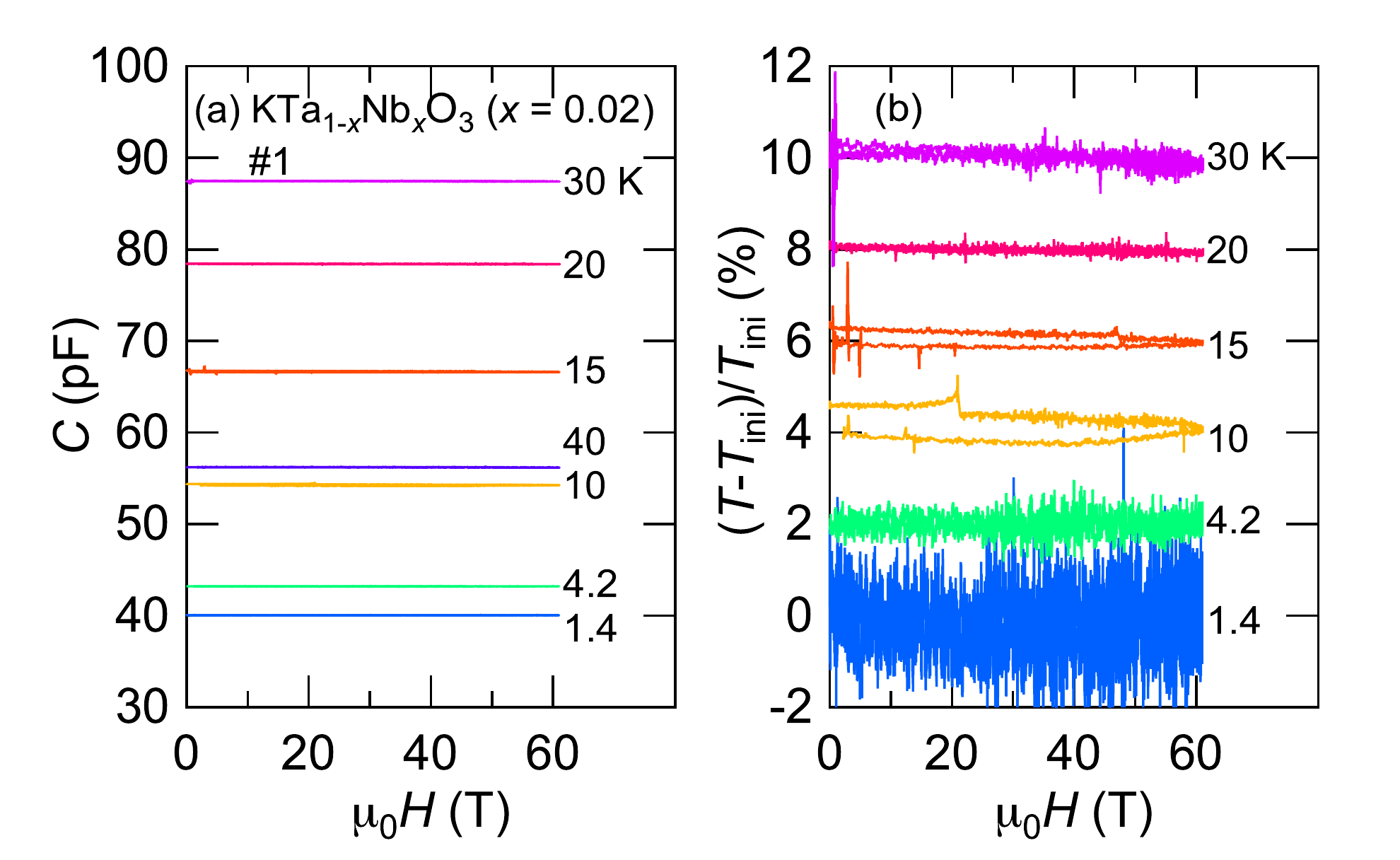}
\caption{(a) Magnetic field dependence of capacitance of KTa$_{1-x}$Nb$_x$O$_3$ ($x$~=~0.02) \#1 at various temperatures. 
(b) Magnetic field dependence of the uncertainty $(T-T_{\rm ini})/T_{\rm ini}$.
For clarify, the data are offset by 2\%.
}
\label{Sd}
\end{center}
\end{figure}

We note here that the $C (T)$ curve varies once the temperature increases across $T_c$, as shown by lines in Fig.~\ref{CH}(a).
This irreproducibility may arise from the nucleation of the ferroelectric domains across $T_c$.
By contrast, the $C(T)$ curve remains reproducible below $T_c$.
We performed heat-cycle test below $T_c$ cooling from room temperature after the $C(H)$ measurements (Fig.~\ref{Sd}).
As shown by colored lines in Fig.~\ref{CH}(a), we can hardly identify the difference in the $C(T)$ curves up to the third cycle.
Therefore, we can safely use the calibration curve of the thermometer unless the temperature of the thermometer does not exceed $T_c$.

Next, we demonstrate the magnetocapacitance effect in KTN.
Figure~\ref{Sd}(a) shows magnetic-field dependences of capacitance up to 61~T at several temperatures.
As expected for nonmagnetic ferroelectrics, KTN shows little magnetic field dependence of capacitance at least below 40~K.
This field dependence is negligibly smaller compared to the temperature dependence, as will be discussed next.

Figure~\ref{Sd}(b) shows the magnetic field variation of $(T-T_{\rm ini})/T_{\rm ini}$, where $T_{\rm ini}$ is the temperature just before applying magnetic field.
Here, we evaluate the temperature using the $C(T)$ curve at zero field.
Although the uncertainty varies with temperature, it is approximately within $\pm$1\%.

We emphasize the advantages of capacitance thermometer as follows.
First, only the temperature scan of capacitance at zero-field is required to use the thermometer in high magnetic-fields.
This simple calibration process is a great advantage of our capacitance thermometer compared to the other thermometers.  
Second, we can manufacture the thermometer having desirable size and shape.
This is important for MCE measurements, as discussed in \cite{Kihara_2013}. 
Third, for metallic samples, this thermometer can be attached without electrical insulation to the sample, which sometimes causes problem in the use of a resistance thermometer.
Finally, since current does not flow through the thermometer, Joule-heating does not emerge.

\subsection{\label{MCE_set}MCE measurements in pulsed-magnetic field using a capacitance thermometer}

\begin{figure}[htbp]
\begin{center}
\includegraphics[width=80mm]{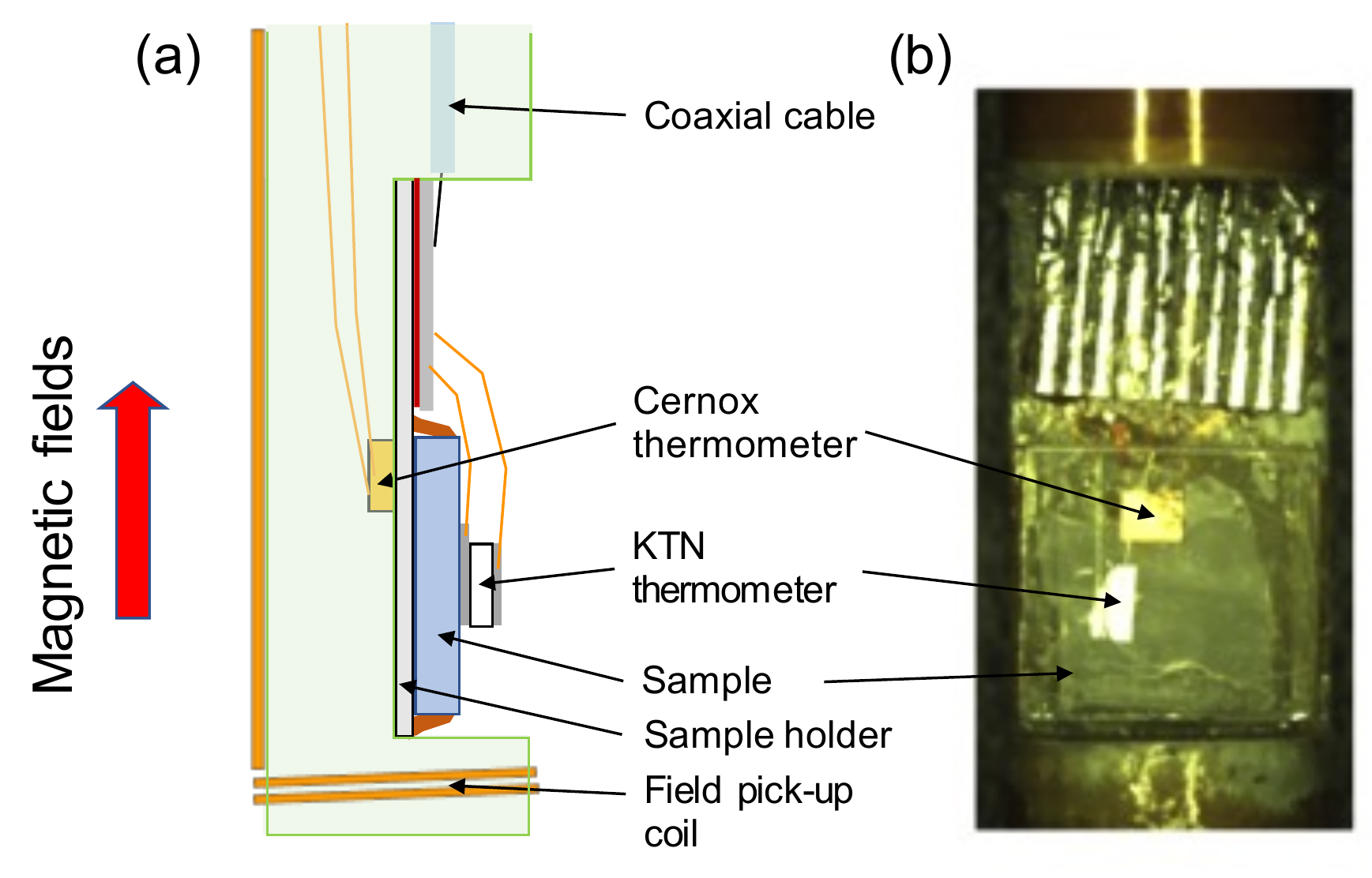}
\caption{(a) Schematic of the probe setup for magneto-caloric effect measurements.
(b) Photograph of an example of the setup.
It is noted that the sample (Gd$_3$Ga$_5$O$_{12}$, here) and the holder made of a quartz plate are transparent, and thus the Cernox{\texttrademark} thermometer behind the holder can be seen. }
\label{setup}
\end{center}
\end{figure}

The setup for the MCE measurement is exemplified in Fig.~\ref{setup}.
To calibrate the capacitance thermometer at zero field, we placeed a calibrated Cernox{\texttrademark} thermometer behind the sample holder made of quartz. 
The sample, Gd$_3$Ga$_5$O$_{12}$ abbreviated as GGG, is glued to the holder by varnish.
KTN thermometer is attached to the top of sample by silver paint, which also works as an electrode of capacitance measurement.
On the opposite side of KTN, another electrode is also made of silver paint.

To keep the sample in quasi adiabatic condition during the MCE measurement, we introduced this probe shown in Fig.~\ref{setup} into a closed tube made of fiber-reinforced plastic. 
The interior of the tube was filled with He gas during the calibration process of the KTN thermometer to isothermalize the KTN and  Cernox{\texttrademark} thermometers.
After the calibration, we evacuated inside of the tube for thermal isolation in the MCE measurement.

\subsection{\label{MCE_GGG}Example of MCE measurement: the case for Gd$_3$Ga$_5$O$_{12}$}

\begin{figure}[htbp]
\begin{center}
\includegraphics[width=\hsize]{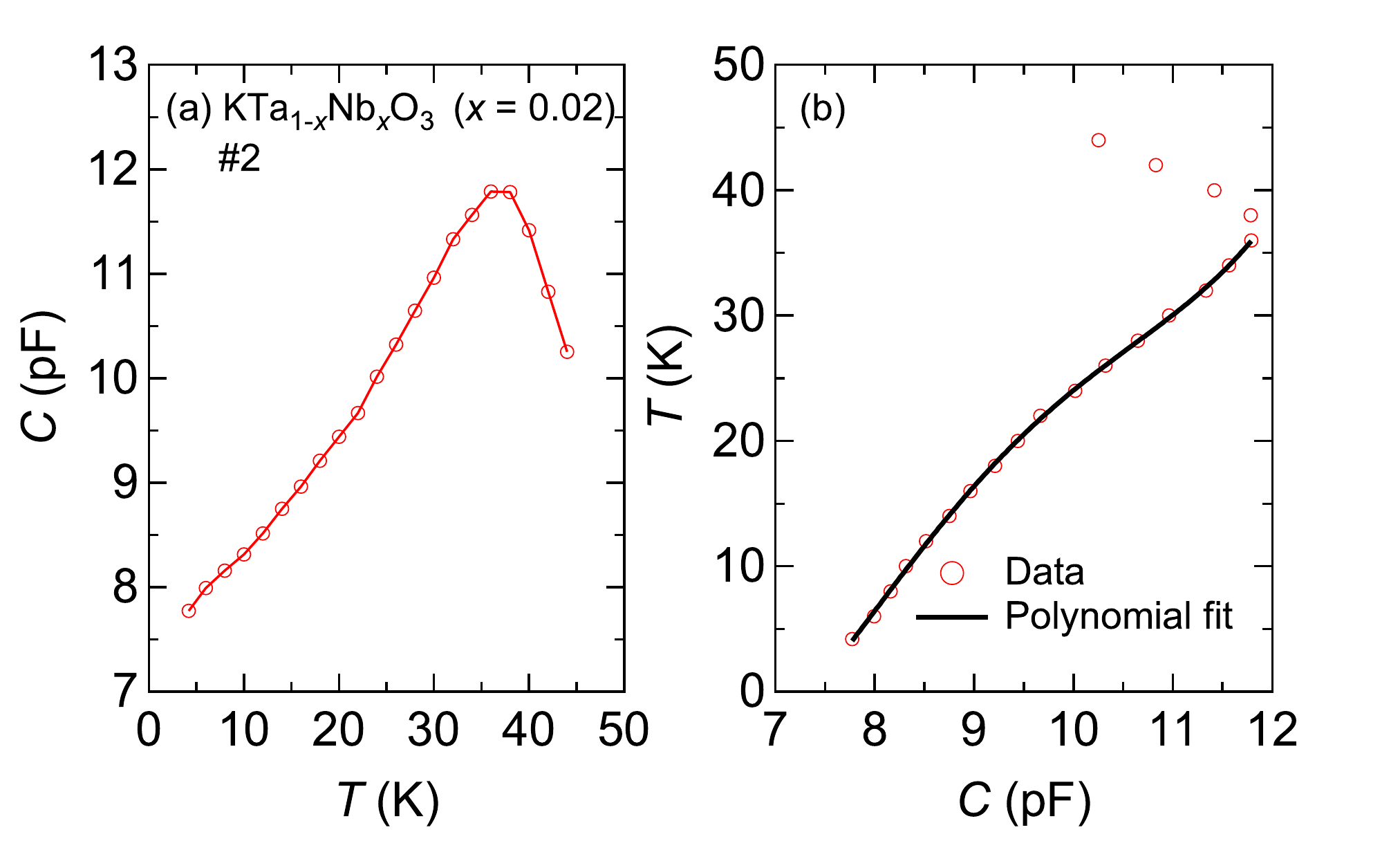}
\caption{(a) Temperature dependence of capacitance of KTa$_{1-x}$Nb$_x$O$_3$ ($x$~=~0.02) \#2 at 0 T measured at 50~kHz. 
The solid line is just for guide.
(b) The relation between capacitance and temperature with a 5th order polynomial fitting result.}
\label{CT}
\end{center}
\end{figure}

We shaped another KTN (\#2) into a thin plate (approximately $0.5 \times 1.4 \times 0.1 {\rm mm}^3$) to reduce the heat capacity and to increase thermal conduction to the sample to realize adiabatic condition in the MCE measurement \cite{Kihara_2013}.
The KTN \#1 and \#2 are taken from the same piece, whose magnetocapacitance effect is confirmed to be as comparable to that of \#1.
Figure~\ref{CT} (a) shows the temperature dependence of capacitance of KTN \#2 attached to GGG.
Below the peak temperature of $\sim$~36~K, capacitance monotonically changes as a function of temperature, which can be expressed by a polynomial curve shown in Fig.~\ref{CT} (b). 
This expression is adopted for the temperature measurements in pulsed-magnetic field.
The difference in peak temperature and the slope of $C(T)$ between KTN \#1 and \#2 may be caused by slight inhomogeneity of chemical composition in the KTN. 

Using this KTN \#2 thermometer, we demonstrate the MCE measurement of GGG.
GGG is a garnet type compound having strong geometric frustration of Gd$^{3+}$ spins, and hence, remains paramagnetic down to 0.4 K \cite{Schiffer_1994}.
It is also confirmed that no magnetic-field-induced phase transition occurs at least up to 40~T above 4.2~K \cite{Levitin_1997}.
In the paramagnetic state, GGG is known to show a large MCE at low temperatures, owing to the large angular momentum $S = 7/2$ for Gd$^{3+}$ \cite{Levitin_1997, Kihara_2013}.
The single crystalline GGG sample was produced by Furuuchi Chemical Corporation.
This sample is the same piece as that used in Ref.~\cite{Kihara_2013}.

\begin{figure}[htbp]
\begin{center}
\includegraphics[width=\hsize]{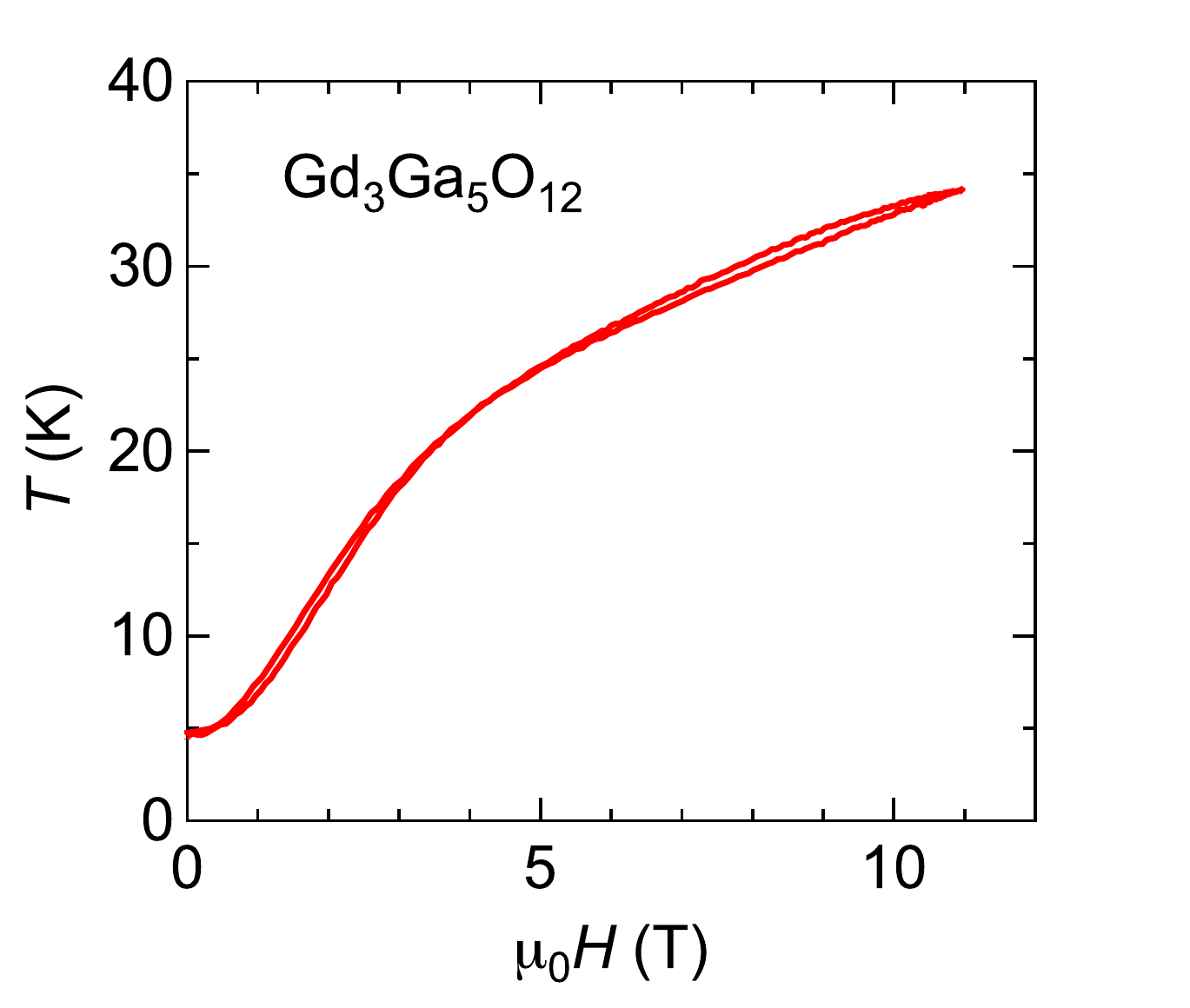}
\caption{Magneto-caloric effect on Gd$_3$Ga$_5$O$_{12}$ measured in pulsed magnetic field using a capacitance thermometer.}
\label{GGG_MCE}
\end{center}
\end{figure}

Figure~\ref{GGG_MCE} shows the magnetic-field dependence of the temperature, namely MCE, of GGG.
The initial temperature set to be approximately 5~K increases with the magnetic field.
At the maximum applied field of 11~T, the temperature reaches 34~K. 
This result is in good agreement with previous results \cite{Levitin_1997, Kihara_2013}.
Since our thermometer used in this study shows a maximum in capacitance at 36~K, the calibration above this temperature does not work well (Fig.~\ref{CT}).
In addition, thermal cycle across the peak temperature sometimes changes the $C(T)$ curve as mentioned previously.
These are reasons why we limit this experiment up to 11~T here.
We can measure the MCE up to 61~T, if the effect is not as large as that in GGG.

\section{Conclusions}

We have developed a system of capacitance measurement in pulsed high magnetic fields up to 61~T.
Using this system, we developed miniature non-metallic capacitance dilatometer cells for magnetostriction measurements.
The efficiency was confirmed by measuring longitudinal magnetostriction in bismuth.
We also measured complex dielectric constants of multiferroic material.
Measurement of the complex dielectric constant could be a powerful tool for a magnetic insulators.
By measuring capacitance of non-magnetic KTa$_{1-x}$Nb$_x$O$_3$ ($x=$0.02), we realized field-insensitive measurement of temperature. 
Using this thermometer, we demonstrated the magneto-caloric effect of Gd$_3$Ga$_5$O$_{12}$ in a pulsed magnetic field.
Our capacitance measuring system is a promising tool to study various magnetic-field-induced phenomena which have been difficult to detect in pulsed magnetic fields. 

\begin{acknowledgments}

We thank W. Knafo for fruitful discussions about magnetostriction measurements.
This work was partially supported by JSPS KAKENHI Grant Numbers JP17H01143, JP19H05823, JP19K21841, JP19H01847, and JP20K03854, and by the MEXT Leading Initiative for Excellent Young Researchers (LEADER).

\end{acknowledgments}

\section*{DATA AVAILABILITY}
The data that support the findings of this study are available from the corresponding author upon reasonable request.

\appendix


\thebibliography{aipsamp}

\bibitem{Herlach_1999} F. Herlach, Rep. Prog. Phys. {\bf 62}, 859 (1999).

\bibitem{White_1961} G. K. White, Cryogenics {\bf 1}, 151 (1961).

\bibitem{Griessen_1973} R. Griessen, Cryogenics {\bf 13}, 375 (1973).

\bibitem{Brooks_1987} J. S. Brooks, M. J. Naughton, Y. P. Ma, P. M. Chaikin, R. V. Chamberlin {\bf 58}, 117 (1987).

\bibitem{Machel_1995} G. Machel, and M. yon Ortenberg, Physica B {\bf 211}, 355 (1995).

\bibitem{Ricodeau_1972} J. A. Ricodeau, D. Melville, and E. W. Lee, J. Phys. E: Sci. Instrum. {\bf 5}, 472 (1972).

\bibitem{Knafo_PC} W. Knafo, private communication.

\bibitem{Levitin_1992} R. Z. Levitin, V. N. Milov, Y. F. Popov, and V. V. Snegirev, Physica B {\bf 177}, 59 (1992).

\bibitem{Ding_2018} X. Ding, Y.-S. Chai, F. Balakirev, M. Jaime, H. T. Yi, S.-W. Cheong, Y. Sun, and V. Zapf, Rev. Sci. Instrum. {\bf 89}, 085109 (2018).

\bibitem{Daou_2010} R. Daou, F. Weickert, M. Nicklas, F. Steglich, A. Haase, and M. Doerr, Rev. Sci. Instru. {\bf 81}, 033909 (2010).

\bibitem{Ikeda_2017} A. Ikeda, T. Nomura, Y. H. Matsuda, S. Tani, Y. Kobayashi, H. Watanabe, and K. Sato, Rev. Sci. Instru. {\bf 88}, 083906 (2017).

\bibitem{Jaime_2017} M. Jaime, C. C. Moya, F. Weickert, V. Zapf, F. F. Balakirev, M. R. Wartenbe, P. F. S. Rosa, J. B. Betts, G. Rodriguez, S. A. Crooker, and R. Daou, Sensors {\bf 17}, 2572 (2017).


\bibitem{Ikeda_2018} A. Ikeda, Y. H. Matsuda, and H. Tsuda, Rev. Sci. Instru. {\bf 89}, 096103 (2018).

\bibitem{Jaime_2010} M. Jaime, R. Daou, S. A. Crooker, F. Weickert, A. Uchida, A. E. Feiguin, C. D. Batista, H. A. Dabkowska, and B. D. Gaulin, Proc. Natl. Acad. Sci. USA {\bf 109}, 12404 (2015).

\bibitem{Rotter_2014} M. Rotter, Z.-S. Wang, A. T. Boothroyd, D. Prabhakaran, A. Tanaka, and M. Doerr, Sci. Rep {\bf 4}, 7003 (2014).

\bibitem{Radtke_2015} G. Radtke, A. Sa$\acute{\rm u}$l, H. A. Dabkowska, M. B. Salamon, and M. Jaime, Proc. Natl. Acad. Sci. USA {\bf 112} 1971 (2015).

\bibitem{Kido_1989} G. Kido, Physica B {\bf 155}, 199 (1989).

\bibitem{Doerr_2008} M. Doerr,  W. Lorenz, T. Neupert, M. Loewenhaupt, N. V. Kozlova, J. Freudenberger, M. Bartkowiak, E. Kampert, and M. Rotter, Rev. Sci. Instrum. {\bf 79}, 063902 (2008).

\bibitem{Kawachi_2017} S. Kawachi, A. Miyake, T. Ito, S. E. Dissanayake, M. Matsuda, W. Ratcliff II, Z. Xu, Y. Zhao, S. Miyahara, N. Furukawa, and M. Tokunaga, Phys. Rev. Materials {\bf 1}, 024408 (2017). 

\bibitem{Lawless_1979} W. N. Lawless, Cryogenics {\bf 19}, 585 (1979).

\bibitem{White_1973} G. K. White, J. Phys. D: Appl. Phys. {\bf 6}, 2070  (1973).

\bibitem{Bunton_1969} G. V. Bunton and S. Weintroub, J. Phys. C: Solid State Phys. {\bf 2}, 116 (1969).

\bibitem{White_1974} G. K. White, AIP Conf. Proc. {\bf 17}, 1 (1974).

\bibitem{Lacerda_1989} A. Lacerda, A. de Visser, P. Haen, P. Lejay, and J. Flouquet, Phys. Rev. B {\bf 40}, 8759 (1989).

\bibitem{Baron_1999} T. H. K. Barron and G. K. White, {\it Heat Capacity and Thermal Expansion at Low Temperatures} (Springer US, Boston, MA, 1999).

\bibitem{Michenaud_1982} J.-P. Michenaud, J. Heremans, M. Shayegan, and C. Haumont, Phys. Rev. B {\bf 26}, 2552 (1982).

\bibitem{Iye_1983} Y. Iye, J. Heremans, K. Nakamura, G. Kido, N. Miura, J.-P. Micheaud, and S. Tanuma, J. Phys. Soc. Jpn. {\bf 52}, 1692 (1983).

\bibitem{Fuseya_2015} Y. Fuseya, M. Ogata, and H. Fukuyama, J. Phys. Soc. Jpn. {\bf 84},
012001 (2015).

\bibitem{Zhu_2017} Z. Zhu, J. Wang, H. Zuo, B. Fauqu$\acute{\rm e}$, R. D. McDonald, Y. Fuseya, and K. Behnia, Nat. Commun., {\bf 8}, 15297 (2017). 

\bibitem{Zhu_2012} Z. Zhu, B. Fauqu$\acute{\rm e}$, L. Malone, A. B. Antunes, Y. Fuseya, and K. Behnia, Proc. Natl Acad. Sci. USA {\bf 109}, 14813 (2012).

\bibitem{Iwasa_2019} A. Iwasa, A. Kondo, S. Kawachi, K. Akiba, Y. Nakanishi, M. Yoshizawa, M. Tokunaga, and K. Kindo, Sci. Rep. {\bf 9}, 1672 (2019). 

\bibitem{Tokura_2014} Y. Tokura, S. Seki, and N. Nagaosa, Rep. Prog. Phys. {\bf 77}, 076501 (2014).

\bibitem{Popov_1993} Y. F. Popov, A. K. Zvezdin, G. P. Vorob'ev, A. M. Kadomtseva, V. A. Murashev, and D. N. Rakov, JETP Lett. {\bf 57}, 69 (1993).




\bibitem{Zapf_2010} V. S. Zapf, M. Kenzelmann, F. Wolff-Fabris, F. Balakirev, and Y. Chen, Phys. Rev. B {\bf 82}, 060402(R) (2010).

\bibitem{Chen_2019} R. Chen, J. F. Wang, Z. W. Ouyang, M. Tokunaga, A. Y. Luo, L. Lin, J. M. Liu, Y. Xiao, A. Miyake, Y. Kohama, C. L. Lu, M. Yang, Z. C. Xia, K. Kindo, and L. Li, Phys. Rev. B {\bf 100}, 140403(R) (2019).


\bibitem{Kimura_2018B} K. Kimura, M. Toyoda, P. Babkevich, K. Yamauchi, M. Sera, V. Nassif, H. M. R{\o}nnow, and T. Kimura, Phys. Rev. B {\bf 97}, 134418 (2018).

\bibitem{Kimura_2018} K. Kimura, Y. Kato, K. Yamauchi, A. Miyake, M. Tokunaga, A. Matsuo, K. Kindo, M. Akaki, M. Hagiwara, S. Kimura, M. Toyoda, Y. Motome, and T. Kimura,
Phys. Rev. Mater. {\bf 2}, 104415 (2018).

\bibitem{Kimura_2019} K. Kimura, S. Kimura, T. Kimura, J. Phys. Soc. Jpn. {\bf 88}, 093707 (2019).


 

\bibitem{Gschneidner_2005} K. A. Gschneidner Jr, V. K. Pecharsky, and A. O. Tsokol, Rep. Prog. Phys. {\bf 68}, 1479 (2005).  

\bibitem{Zhu_2003} L. Zhu, M. Garst, A. Rosch, and Q. Si, Phys. Rev. Lett. {\bf 91}, 066404 (2003).

\bibitem{Garst_2005} M. Garst and A. Rosch, Phys. Rev. B {\bf 
72}, 205129 (2005).

\bibitem{Levitin_1997} R. Z. Levitin, V. V. Snegirev, A. V. Kopylov, A. S. Lagutin, and A. Gerber, J. Magn. Magn. Mater. {\bf 170}, 223 (1997).

\bibitem{Dankov_1997} S. Y. Dan'kov, A. M. Tishin, V. K. Pecharsky, and K. A. Gschneidner, Jr., Rev. Sci. Instrum. {\bf 68}, 2432 (1997).

\bibitem{Kohama_2010} Y. Kohama, C. Marcenat, T. Klein, and M. Jaime, Rev. Sci.
Instrum. {\bf 81}, 104902 (2010).

\bibitem{Kihara_2013} T. Kihara, Y. Kohama, Y. Hashimoto, S. Katsumoto, and M. Tokunaga, Rev. Sci. Instrum. {\bf 84}, 074901 (2013).

\bibitem{Kohama_2013} Y. Kohama, Y. Hashimoto, S. Katsumoto, M. Tokunaga, and K. Kindo, Meas. Sci. Technol. {\bf 24}, 115005 (2013).

\bibitem{Gottschall_2016} T. Gottschall, K. P. Skokov, F. Scheibel, M. Acet, M. G. Zavareh, Y. Skourski, J. Wosnitza, M. Farle, and O. Gutfleisch, Phys. Rev. Applied {\bf 5}, 024013 (2016).

\bibitem{Brandt_1999} B. L. Brandt, D. W. Liu, L. G. Rubin, Rev. Sci. Instrum. {\bf 70}, 104 (1999).

\bibitem{Scott_2003}S. S. Courts, and Philip R. Swinehart, AIP Conf. Proc. {\bf 684}, 393 (2003).

\bibitem{Rosenbaum_2001} R. Rosenbaum, B. Brandt, S. Hannahs, T. Murphy, E. Palm, B. J. Pullum, Physica B {\bf 294-295}, 489 (2001).

\bibitem{Murphy_2001} T. P. Murphy, E. C. Palm, L. Peabody, and S. W. Tozer, Rev. Sci. Instum. {\bf 72}, 3462 (2001).

\bibitem{Tetsuka_2006} H. Tetsuka, H. Takashima, B. Prijamboedi, R. Wang, A. Shoji, Y. J. Shan, M. Itoh, Phs. Stat. Sol. (a) {\bf 203}, 2546 (2006).

\bibitem{Shimada_2006} K. Shimada, H. Takashima, R. Wang, B. Prijamboedi, N. Miura, M. Itoh, Ferroelectrics {\bf 331}, 141 (2006).
 
\bibitem{Naughton_1983} M. J. Naughton, S. Dickinson, R. C. Samaratunga, J. S. Brooks, and K. P. Martin, Rev. Sci. Instrum. {\bf 54}, 1529 (1983).

\bibitem{Rytz_1983} D. R. Rytz, A. Ch$\hat{\rm a}$telain, and U. T. H$\ddot{\rm o}$chli, Phys. Rev. B {\bf 27}, 6830 (1983). 

\bibitem{Tachibana_2015} M. Tachibana, Solid State Commun. {\bf 221}, 33 (2015). 

\bibitem{Schiffer_1994} P. Schiffer, A. P. Ramirez, D. A. Huse, and A. J. Valentino, Phys. Rev. Lett. {\bf 73}, 2500 (1994).
\end{document}